# Crossover behavior in the magnetoresistance of thin flakes of the topological material ZrTe$_5$

*Zhijian Xie,[1,2] Xinjian Wei,[1,3] Xiaobin Qiang,[4] Yu Zhang,[3] Shili Yan,[3] Shimin Cao,[1,3] Congkuan Tian, [1,3] Peipei Wang,[4] Liyuan Zhang,[4] G. D. Gu,[5] Haizhou Lu,[4] Jian-Hao Chen[1,2,3,6]*

[1]*International Center for Quantum Materials, School of Physics, Peking University, Beijing 100871, China*
[2]*Key Laboratory for the Physics and Chemistry of Nanodevices, Peking University, Beijing 100871, China*
[3]*Beijing Academy of Quantum Information Sciences, Beijing 100193, China*
[4]*Department of Physics and Shenzhen Institute for Quantum Science and Engineering, Southern University of Science and Technology, Shenzhen, China*
[5]*Condensed Matter Physics & Materials Science Division, Brookhaven National Laboratory, Upton, New York 11973-5000, USA*
[6]*Interdisciplinary Institute of Light-Element Quantum Materials and Research Center for Light-Element Advanced Materials, Peking University, Beijing 100871, China*

**Abstract**

ZrTe$_5$ is a layered material that exhibits intricate topological effects. Intensive theoretically and experimental efforts have been devoted to try to understand the physics in this materials. In this paper the temperature dependent magneto-transport properties of ZrTe$_5$ thin flakes are investigated. A characteristic temperature $T^*$ is observed in the temperature dependence of three different types of magnetoresistance simultaneously, which are the saturated Hall anomaly, the chiral anomaly and the longitudinal magnetoresistance. Furthermore, the value of $T^*$ decreases monotonically from 200K to 160K with increasing thickness of the ZrTe$_5$ thin flakes from 42nm to 89nm. Temperature induced topological phase transitions are attributed to the cause of such anomaly in the three types of magnetoresistance at $T^*$. Our findings provide a multi-parameter indicator for the emergence of topological phase transition in ZrTe$_5$ and could be extended to the study of other topological materials. The temperature dependence of the three types of magnetoresistance also shed light on the role of anomalous Hall Effect in the transport properties of ZrTe$_5$.

**Introduction**

The anomalous Hall effect (AHE), having a history almost as long as the Hall effects, is frequently observed in ferromagnetic materials in which time reversal symmetry is broken [1]. Tremendous efforts have been devoted to clarify the microscopic physical origin of AHE. Typically, three mechanisms are believed to give rise to AHE: side jump, skew scattering and intrinsic deflection [1]. For a very long time these mechanisms of AHE are restricted in ferromagnetic materials or non-magnetic materials doped with magnetic impurities. In the recent years, progress has been made to understand AHE originating from non-magnetic topological materials such as Dirac semimetals and Weyl semimetals [2-5]. In topological materials, intrinsic contribution to the AHE may stem from the Berry curvature. Specifically, in Weyl semimetals, the Weyl nodes act like magnetic monopoles in the momentum space [2,6]. On the other hand, in a number of Weyl and Dirac semimetals, multiple types of charge carriers at the Fermi level can also lead to nonlinearity in the Hall resistance versus magnetic field curves [7,8], making it difficult to pinpoint the actual physical origin of such effects.

Transition metal pentatelluride $ZrTe_5$ is a fascinating material that is considered to be at the boundary of strong and weak topological insulators, resembling Dirac semimetals [2,9-14]. Monolayer $ZrTe_5$ has been theoretically predicted to be a large gap quantum spin Hall insulator [15]. A number of experiments have confirmed the non-trivial Berry phase and Dirac-like dispersion [12,16] in $ZrTe_5$ through the observation of quantum oscillation [13,14,17,18], chiral anomaly induced negative magnetoresistance [19], anomalous Hall effect [2,3], planar Hall effect [20] and three dimensional quantum Hall effect [11], highlighting $ZrTe_5$ as a promising platform to study quantum and topological phenomena. In particular, various experimental attempts have been made to try to induce topological phase transition in $ZrTe_5$, including physical variables such as strain [9], temperature [21,22] and hydrostatic pressure [23].

In this work we measure the transport properties of thin flakes of $ZrTe_5$ with perpendicular magnetic field and with in-plane magnetic field parallel with the current direction at temperatures $T$ ranging from 2K to 300K. We found that a single characteristic temperature $T^*$ exists for any particular $ZrTe_5$ thin flake, at which crossover behavior appears simultaneously in the $T$ dependence of three types of magnetoresistance: 1) the saturated Hall anomaly under perpendicular magnetic field ($R_{AH}^S(B_\perp)\ vs.T$), 2) the chiral anomaly under in-plane magnetic field along the current direction ( $NMR(B_\parallel)\ vs.T$ ), and 3) the longitudinal magnetoresistance under perpendicular magnetic field ($MR(B_\perp)\ vs.T$).

First, the crossover behavior in $NMR(B_\parallel)\ vs.T$ can be attributed to the intrinsic effect of a divergent Berry curvature in the material [2,3]. Second, the appearance of the crossover behavior in multiple transport properties at a single $T^*$ strongly suggests that such temperature marks the initiation of the temperature-induced topological phase transition in ZrTe$_5$ [12,24]. Furthermore, $T^*$ decreases monotonically with increasing thickness of the ZrTe$_5$ thin flakes, suggesting that such topological phase transition is highly tunable by sample thickness. Our results also indicate that the observed nonlinearity of the Hall curve contains the contribution of Berry curvature as well as multi bands effect.

**Results**

Figure 1a shows the atomic schematics of the crystal structure of ZrTe$_5$. Zirconium atoms bond along the $a$ axis, and each zirconium atom is related to three tellurium atoms, forming ZrTe$_3$ chains. The ZrTe$_3$ chains are linked by two tellurium atoms along the $c$ axis. Then the layered structure stacks along the $b$ axis by van der Waals interactions. This peculiar structure makes ZrTe$_5$ easily exfoliated. Figure 1b is the 3D micrograph of a typical ZrTe$_5$ device taken by an atomic force microscope (AFM). Figure 1c shows the temperature dependent resistance curve of the device, featured by a resistance peak at $T_p = 123K$. The resistance peak is widely observed in ZrTe$_5$ and can be tuned by crystal quality, flake thickness and magnetic field [10,11,19]. The position of $T_p$ is associated with the Lifshitz transition in ZrTe$_5$, at which the Fermi level shifts from the conduction band to the valence band [11,12] and accompanied by sign reversal of the Hall coefficient [11,25,26]. The inset in figure 1c denotes the axes related to the sample crystal direction in the device shown in figure 1b, as well as the direction of the current. The thickness of the channel is 42 nm as determined by AFM. Since other devices show similar behavior, we shall focus our discussion on the data obtained from device 1.

Figure 2a shows the magneto-resistance ( $MR = \frac{R(B)-R(B=0)}{R(B=0)} \times 100\%$ ) at different temperatures, large and positive $MR$ is observed at all temperatures, similar to previous report in this material and other topological materials, e.g., WTe$_2$ and Cd$_3$As$_2$ [7,8]. Such large $MR$ is a manifestation of carrier compensation from different bands [27], and the electron scattering in ZrTe$_5$ can be indirectly probed using the Kohler's plot (Figure 2b). The Kohler's plot, i.e. plotting *MR* vs $B^2/R_{xx}^2(B=0)$, is a convenient way to visualize whether or not different types of scattering appear in a particular sample at different temperatures. If there exists only one type of scattering, *MR* of a sample at different temperatures should conform to the same functional form:

$$MR = F(B^2/R_{xx}^2(B=0)) \qquad \text{Eq. 1}$$

Thus in this case, all the curves in the Kohler's plot would fall into one single trace. As can be seen in Figure 2b, at low temperatures (2-160K), *MR* curves in the Kohler's plot stay close to each other. While the temperature rises above ~160K, on the other hand, the *MR* curves shift systematically. This implies that with increasing temperature, more scattering mechanism sets in, possibly due to the crossing of the Fermi level with a new band. Figure 2c shows the Hall resistance at temperatures ranging from 2K to 300K, in which nonlinear Hall curves are observed. The Hall resistance can be separated into two parts: $R_{xy} = R_n + R_{AH}$, where the first part $R_n$ is the contribution from normal Hall resistance arising from the Lorentz force. The second part $R_{AH}$ denotes the Hall anomaly, which is shown in Figure 2d. The Hall anomaly is considered to be the result of anomalous Hall effect in non-magnetic material ZrTe$_5$ [2-5]. Although a conclusive understanding of this anomalous Hall effect is still lacking, the community has reached a consensus that Berry curvature plays a significant role. We note that the sign reversal of the Hall coefficient is missing in thin devices studied in this work, which is in contrast to thicker device (thickness more than 170 nm, see Supplemental Materials S1 for more details [33]) and bulk crystals reported previously [11,25,26]. This may be a result of band structure evolution with sample thickness and warrants further study in the future.

Next, we rotate the magnetic field from the ***b*** axis to the ***a*** axis. In this case the magnetic field is parallel to the direction of the current. In this configuration, we measure the magnetoresistance, as shown in Figure 3a. Negative *MR* (*NMR*) with parallel external electric and magnetic field is believed to be a strong signature of chiral anomaly in Weyl semimetals such as ZrTe$_5$ and TaAs [19,28]. Such *NMR* is a result of the topological ***E · B*** term that gives rise to unbalanced carrier density with opposite chirality. Magneto-conductivity induced by chiral anomaly can be described by [19]:

$$\sigma = \sigma_0 + C_w \cdot B^2 \qquad \text{Eq. 2}$$

where $C_W = \frac{3}{8}\frac{e^5}{\pi^4\hbar^2 c}\frac{v^3\tau_V}{T^2+(\mu/\pi)^2}$ [6], in which $e$ is the elementary charge, $\hbar$ is the reduced Plank constant, $\tau_v$ is the chirality-changing scattering time, $v$ is Fermi velocity, $\mu$ is the chemical potential and $c$ is the speed of light [19]. The magneto-conductivity $\sigma$ at different temperatures (converted from figure 3a) is fitted by Eq. 2. The coefficient $C_W$ decreases with increasing temperature, as shown in Figure 3b. $C_W$ is well-fitted to Eq. 2 and it is nonzero up to a temperature of ~190K.

From the data mentioned above, a crossover behavior is found at three different physical quantities versus temperature: 1) the saturated Hall anomaly with a large enough perpendicular magnetic field $R_{AH}^s(B_\perp = 9T)$; 2) the amplitude of *NMR* with a magnetic field of 9T along the current direction; 3) magneto-resistance with perpendicular magnetic field $MR(B_\perp = 9T)$. Furthermore, highly similar results are also observed for other ZrTe$_5$ devices with thickness ranging from 42nm to 89nm, with intriguing and systematic changes. As an example, these three quantities in three ZrTe$_5$ devices with different thickness (42 nm, 54 nm and 89 nm, respectively) are plotted in Figure 4. The upper panel in Figure 4 shows $R_{AH}^s(B_\perp = 9T)$ versus temperature; the middle panel shows *NMR vs. T*; and the lower shows $MR(B_\perp = 9T)$ *vs. T*. One can see that there is a crossover for all three quantities in device 1 with a characteristic temperature $T^* \approx 200K$. For device 2 and device 3, $T^* \approx 180K$ and 160K, respectively. Here we choose the temperature corresponding to maximum *MR* as the $T^*$. Interestingly, it can be seen that the thinner the sample, the higher the value of $T^*$. We plot $T^*$ from six different samples exfoliated from two different batches of bulk ZrTe$_5$ crystals in Figure 5, which show the monotonic trend clearly.

In order to understand the crossover behavior, e.g., why a single $T^*$ appears in three physical quantities in one device and why $T^*$ changes monotonically with a changing thickness of the ZrTe$_5$ crystals, we look into the physical origins of the Hall anomaly. Generally, there are three physical origins of the Hall anomaly: 1) the intrinsic mechanism related to Berry curvature; 2) magnetic scatterings (including side jump and skew scattering); 3) multiband effect. Since ZrTe$_5$ has no magnetic element and no experiment demonstrates magnetism in this material, we can rule out the possibility of magnetic scatterings. The *NMR* at **B** ∥ **I** combined with the $R_{AH}^s$ under **B**$_\perp$ indicates that they likely originated from Berry curvature generated by the Weyl nodes [2]. Weyl nodes act like magnetic monopoles, generating diverging Berry curvature directed radially inward or outward [6]. For any specific ZrTe$_5$ device, the temperature dependence of *NMR* and $R_{AH}^s$ all show a crossover at $T = T^*$ indicate that the Berry curvature plays an important role in such behavior. Below $T^*$, the chiral anomaly induced by the Weyl nodes leads to the *NMR*, and the related Berry curvature gives rise to the *AHE*, which is more of the two sides of the same stone. The fact that *NMR* disappears above $T^*$ while *MR* and *AHE* remains finite at temperature ranging from 2K to 300K (albeit a kink appears at $T^*$) provides us with a picture that Weyl nodes and Berry curvature start to make contributions below $T^*$, and multiband effect is likely to be in effect throughout the temperature range we explored experimentally.

To confirm this hypothesis, we performed theoretical calculations (see Supplemental Material S3 for more details [33]) on the Berry curvature induced, temperature dependent anomalous Hall conductivity under $\boldsymbol{B}_\perp$ and negative magnetoresistance with $\boldsymbol{B} \parallel \boldsymbol{I}$. The calculation is based on Boltzmann transport theory, with its applicability widely tested by experiments in topological materials [29]. For ZrTe$_5$ we used the effective Hamiltonian [30,31] to explore the transport properties:

$$H_0 = \hbar(v_x k_x \tau_x \otimes \sigma_z + v_y k_y \tau_y \otimes \sigma_0 + v_z k_z \tau_x \otimes \sigma_x)$$
$$+[M_0 + M_1(v_x^2 k_x^2 + v_y^2 k_y^2) + M_z k_z^2]\tau_z \otimes \sigma_0 \qquad \text{Eq. 3}$$

The model parameters are $v_x = 9\times10^5$ m/s, $v_y = 1.9\times10^5$ m/s; $v_z = 0.3\times10^5$ m/s, $M_0 = -4.7$ meV, $M_1 = 150$ meV·nm$^2$, $M_z = 0.01M_1$. With non-zero magnetic field, an additional Zeeman term needs to be considered:

$$H_z = -\frac{\mu_B}{2}(\bar{g}\tau_0 \otimes \boldsymbol{\sigma} + \delta g \tau_z \otimes \boldsymbol{\sigma}) \cdot \boldsymbol{B} \qquad \text{Eq. 4}$$

where $\bar{g} = 22.5$, $\delta g = 11.25$ are $g$ factors, and $\mu_B$ is the Bohr magneton. The complete Hamiltonian is thus $H = H_0 + H_z$.

The effect of magnetic field comes from two different sides. First, magnetic field induced Zeeman splitting breaks the time reversal symmetry and leads to the non-zero Berry curvature. In the linear response region, the intrinsic anomalous Hall effect can be directly associated to the Berry curvature [32]. Second, coupling between magnetic field and Berry curvature gives rise to the anomalous velocity to the Bloch electrons, which is the main ingredient leading to the negative magnetoresistance. As shown in figure 6, 1) both the calculated anomalous Hall conductivity under $\boldsymbol{B}_\perp$ and negative magnetoresistance with $\boldsymbol{B} \parallel \boldsymbol{I}$ trend to 0 with the increase of temperature; 2) both quantities get close to the minimum value at the same temperature (around 200 K). The above two features agrees very well with the experimental result, showing that such crossover behavior indeed comes from Berry curvature effects. Note that the Hamiltonian we have used is a 3D Dirac model, thus the results reveal the properties of bulk material in principle. In our calculation we phenomenologically suppose that the chemical potential varies slowly with temperature from the Fermi energy, which might be complicated in real materials. Subtle variation between the real material and the theoretical model could have contributed to a higher value of the calculated $T^*$ as compares with experimental ones.

Experimentally, similar observation has been made in $WTe_{1.98}$ crystal, in which the vanishing *NMR* is attributed to the temperature induced topological phase transition from Weyl semimetal to a trivial semimetal [24]. Furthermore, signatures of temperature induced topological phase transition in $ZrTe_5$ are demonstrated by infrared spectroscopy and thermodynamic study [21,22]. Since Hall anomaly and positive *MR* are widely observed in topological materials such as $Cd_3As_2$ and $TaAs$, our results may provide a general indicator as of when topological effects kick in.

**Conclusion**

In summary, we have observed a crossover behavior in the transport properties of $ZrTe_5$ thin flakes with thickness ranging from 42nm to 89nm. At the characteristic temperature $T^*$, longitudinal magnetoresistance and Hall anomaly with perpendicular magnetic field as well as negative magnetoresistance arising from chiral anomaly all show a crossover, which we attribute to a topological phase transition induced by temperature. The characteristic temperature evolves from 160K to 200K as sample thickness reduces from 89nm to 42nm. Our finding could provide a multi-parameter indicator for the emergence of topological phase transition in $ZrTe_5$ and other topological materials. The temperature dependence of Hall curves suggests that the both Berry curvature and multi types carrier contribute to the nonlinear Hall curve.

**Methods**

The $ZrTe_5$ flakes are mechanically exfoliated on silicon substrate with 285nm oxide. Standard e-beam lithography was used to pattern electrodes followed by an e-beam evaporation of Cr (5 nm) and Au (60 nm). The electrical measurements were performed in a physical property measurement system (PPMS) with standard lock-in technique. Special attention has been paid to protect the thin flakes from exposing to ambient conditions. All the sample preparation process and device fabrication process were done in vacuum, inert atmosphere, or with the sample capped with a protection layer. The protection layer consists of a bilayer of 200-nm polymethyl methacrylate (PMMA) and 200-nm methyl methacrylate (MMA).

**Author Contributions**

Z.J.X. and J.-H.C. conceived the experiment. L.Z., P.W. and G.D.G provided high quality bulk $ZrTe_5$ crystals, Z.J.X. fabricated the devices, performed transport measurements, and analyzed the data. X.J.W, S.Y. and S.C. aided in transport measurements, Y.Z. performed atomic force microscopy measurements, H.Z.L. and X.B.Q. performed the theoretical calculations. Z.J.X., and J.-H.C. discussed the results and analyzed the data. Z.J.X. and J.-H.C. wrote the manuscript, C.K.T. participated in revising the manuscript and all authors commented on it.

**Notes**

The authors declare no competing financial interest.

**Acknowledgements**

The authors thank X. C. Xie and S. Q. Shen for helpful discussions. This project has been supported by the National Basic Research Program of China (Grant Nos. 2019YFA0308402, 2018YFA0305604), the National Natural Science Foundation of China (NSFC Grant Nos. 11934001, 11774010, 11921005), Beijing Municipal Natural Science Foundation (Grant No. JQ20002). The work at Brookhaven National Laboratory was supported by the U. S. Department of Energy (DOE), Office of Basic Energy Sciences, Division of Materials Sciences and Engineering, under Contract No. DE-SC0012704.  L.Z. acknowledges support by the National Natural Science Foundation of China ((Grants No. 11874193) and Shenzhen Fundamental Research Fund for Distinguished Young Scholars (Grants No. RCJC2020071414435105).

[†]Corresponding author: chenjianhao@pku.edu.cn

**References**

[1] N. Nagaosa, S. Jairo, O. Shigeki, A. H. Macdonald and N. P. Ong, Anomalous Hall effect, *Rev. Mod. Phys.* **82**, 1539 (2010).

[2] T. Liang, J. J. Lin, Q. Gibson, S. Kushwaha, M. H. Liu, W. D. Wang, H. Y. Xiong, J. A. Sobota, M. Hashimoto, P. S. Kirchmann, Z. X. Shen, R. J. Cava and N. P. Ong, Anomalous Hall effect in $ZrTe_5$, *Nat. Phys.* **14**, 451 (2018).

[3] Z. L. Sun, Z. P. Cao, J. H. Cui, C. S. Zhu, D. H. Ma, H. H. Wang, W. Z. Zhuo, Z. H. Cheng, Z. Y. Wang, X. G. Wan and X. H. Chen, Large Zeeman splitting induced anomalous Hall effect in $ZrTe_5$, *Npj: Quantum Mater*. **5**, 36 (2020).

[4] Joshua Mutch, Xuetao Ma, Chong Wang, Paul Malinowski, Joss Ayres-Sims, Qianni Jiang, Zhoayu Liu, Di Xiao, Matthew Yankowitz and J.-H. Chu, Abrupt switching of the anomalous Hall effect by field-rotation in nonmagnetic ZrTe5, *arxiv*: 2101.02681 (2021).

[5] Yanzhao Liu, Huichao Wang, Huixia Fu, Jun Ge, Yanan Li, Chuanying Xi, Jinglei Zhang, Jiaqiang Yan, David Mandrus, Binghai Yan and J. Wang, Induced anomalous Hall effect of massive Dirac fermions in $ZrTe_5$ and $HfTe_5$ thin flakes, *arxiv*: 2012.08188 (2021).

[6] F. Zhong, N. Naoto, T. Kei S., A. Atsushi, M. Roland, O. Takeshi, Y. Hiroyuki, K. Masashi, T. Yoshinori and T. Kiyoyuki, The anomalous Hall effect and magnetic monopoles in momentum space, *Science*, **302**, 92 (2003).

[7] X. Liu, Z. Zhang, C. Cai, S. Tian, S. Kushwaha, H. Lu, T. Taniguchi, K. Watanabe, R. J. Cava, S. Jia and J.-H. Chen, Gate tunable magneto-resistance of ultra-thin $WTe_2$ devices, *2D Mater.* **4**, 021018 (2017).

[8] C. Z. Li, J. G. Li, L. X. Wang, L. Zhang, J. M. Zhang, D. Yu and Z. M. Liao, Two-carrier transport induced Hall anomaly and large tunable magnetoresistance in Dirac semimetal $Cd_3As_2$ nanoplates, *ACS Nano*, **10**, 6020-6028 (2016).

[9] J. Mutch, W.-C. Chen, P. Went, T. Qian, I. Z. Wilson, A. Andreev, C.-C. Chen and J.-H. Chu, Evidence for a strain-tuned topological phase transition in $ZrTe_5$, *Sci. Adv.* **5**, eaav9771 (2019).

[10] J. Niu, J. Wang, Z. He, C. Zhang, X. Li, T. Cai, X. Ma, S. Jia, D. Yu and X. Wu, Electrical transport in nanothick $ZrTe_5$ sheets: From three to two dimensions, *Phys. Rev. B* **95**, 035420 (2017).

[11] F. Tang, Y. Ren, P. Wang, R. Zhong, J. Schneeloch, S. A. Yang, K. Yang, P. A. Lee, G. Gu, Z. Qiao and L. Zhang, Three-dimensional quantum Hall effect and metal-insulator transition in $ZrTe_5$, *Nature*, **569**, 537-541 (2019).

[12] Y. Zhang, C. Wang, L. Yu, G. Liu, A. Liang, J. Huang, S. Nie, X. Sun, Y. Zhang, B. Shen, J. Liu, H. Weng, L. Zhao, G. Chen, X. Jia, C. Hu, Y. Ding, W. Zhao, Q. Gao, C. Li, S. He, L. Zhao, F. Zhang, S. Zhang, F. Yang, Z. Wang, Q. Peng, X. Dai, Z. Fang, Z. Xu, C. Chen and X. J. Zhou, Electronic evidence of temperature-induced Lifshitz transition and topological nature in $ZrTe_5$, *Nat. Commun.* **8**, 15512 (2017).

[13] J. Wang, J. Niu, B. Yan, X. Li, R. Bi, Y. Yao, D. Yu and X. Wu, Vanishing quantum oscillations in Dirac semimetal $ZrTe_5$, *Proc. Natl. Acad. Sci.* **115**, 9145 (2018).

[14] G. L. Zheng, J. W. Lu, X. D. Zhu, W. Ning, Y. Y. Han, H. W. Zhang, J. L. Zhang, C. Y. Xi, J. Y. Yang, H. F. Du, K. Yang, Y. H. Zhang and M. L. Tian, Transport evidence for the three-dimensional Dirac semimetal phase in $ZrTe_5$, *Phys. Rev. B* **93**, 115414 (2016).

[15] H. M. Weng, X. Dai and Z. Fang, Transition-metal pentatelluride $ZrTe_5$ and $HfTe_5$, A paradigm for large-gap quantum spin Hall insulators, *Phys. Rev. X* **4**, 011002 (2014).

[16] H. Xiong, J. A. Sobota, S. L. Yang, H. Soifer, A. Gauthier, M. H. Lu, Y. Y. Lv, S. H. Yao, D. Lu, M. Hashimoto, P. S. Kirchmann, Y. F. Chen and Z. X. Shen, Three-dimensional nature of the band structure

[17] M. K. Hooda and C. S. Yadav, Unusual magnetoresistance oscillations in preferentially oriented p-type polycrystalline $ZrTe_5$, *Phys. Rev. B* **98**, 165119 (2018).

[18] Y.-Y. Lv, B.-B. Zhang, X. Li, K.-W. Zhang, X.-B. Li, S.-H. Yao, Y. B. Chen, J. Zhou, S.-T. Zhang, M.-H. Lu, S.-C. Li and Y.-F. Chen. Shubnikov–de Haas oscillations in bulk $ZrTe_5$ single crystals: Evidence for a weak topological insulator, *Phys. Rev. B* **97**, 115137 (2018).

[19] Q. Li, D. E. Kharzeev, C. Zhang, Y. Huang, I. Pletikosic, A. V. Fedorov, R. D. Zhong, J. A. Schneeloch, G. D. Gu and T. Valla, Chiral magnetic effect in $ZrTe_5$, *Nat. Phys.* **12**, 550 (2016).

[20] P. Li, C. H. Zhang, J. W. Zhang, Y. Wen and X. X. Zhang, Giant planar Hall effect in the Dirac semimetal $ZrTe_{5-\delta}$, *Phys. Rev. B* **98**, 121108(R) (2018).

[21] N. L. Nair, P. T. Dumitrescu, S. Channa, S. M. Griffin, J. B. Neaton, A. C. Potter and J. G. Analytis, Thermodynamic signature of Dirac electrons across a possible topological transition in $ZrTe_5$, *Phys. Rev. B* **97**, 041111(R) (2018).

[22] B. Xu, L. X. Zhao, P. Marsik, E. Sheveleva, F. Lyzwa, Y. M. Dai, G. F. Chen, X. G. Qiu and C. Bernhard, Temperature-driven topological phase transition and intermediate Dirac semimetal phase in $ZrTe_5$, *Phys. Rev. Lett.* **121**, 187401 (2018).

[23] J. L. Zhang, C. Y. Guo, X. D. Zhu, L. Ma, G. L. Zheng, Y. Q. Wang, L. Pi, Y. Chen, H. Q. Yuan and M. L. Tian, Disruption of the accidental Dirac semimetal state in $ZrTe_5$ under hydrostatic pressure, *Phys. Rev. Lett.* **118**, 206601 (2017).

[24] Y. Y. Lv, X. Li, B. B. Zhang, W. Y. Deng, S. H. Yao, Y. B. Chen, J. Zhou, S. T. Zhang, M. H. Lu, L. Zhang, M. Tian, L. Sheng and Y. F. Chen, Experimental observation of anisotropic Adler-Bell-Jackiw anomaly in type-II Weyl semimetal $WTe_{1.98}$ crystals at the quasiclassical regime, *Phys. Rev. Lett.* **118**, 096603 (2017).

[25] H. Chi, C. Zhang, G. Gu, D. E. Kharzeev, X. Dai and Q. Li. Lifshitz transition mediated electronic transport anomaly in bulk $ZrTe_5$, *New J Phys*, **19**, 015005 (2017).

[26] L. Zhou, A. Ramiere, P. B. Chen, J. Y. Tang, Y. H. Wu, X. Lei, G. P. Guo, J. Q. He and H. T. He, Anisotropic Landau level splitting and Lifshitz transition induced magnetoresistance enhancement in $ZrTe_5$ crystals, *New J Phys*, **21**, 093009 (2019).

[27] M. N. Ali, J. Xiong, S. Flynn, J. Tao, Q. D. Gibson, L. M. Schoop, T. Liang, N. Haldolaarachchige, M. Hirschberger, N. P. Ong and R. J. Cava, Large, non-saturating magnetoresistance in $WTe_2$, *Nature*, **514**, 205 (2014).

[28] X. Huang, L. Zhao, Y. Long, P. Wang, D. Chen, Z. Yang, H. Liang, M. Xue, H. Weng, Z. Fang, X. Dai and G. Chen, Observation of the Chiral-anomaly-induced negative magnetoresistance in 3D Weyl semimetal TaAs, *Phys. Rev. X* **5**, 031023 (2015).

[29] X. Dai, Z. Z. Du and H. Z. Lu, Negative magnetoresistance without Chiral anomaly in topological insulators, *Phys. Rev. Lett.* **119**, 166601 (2017).

[30] F. Qin, S. Li, Z. Z. Du, C. M. Wang, H. Z. Lu and X. C. Xie, Theory for the charge-density-wave mechanism of 3D quantum Hall effect, *phys. Rev. Lett.* **125**, 206601 (2020).

[31] R. Y. Chen, Z. G. Chen, X. Y. Song, J. A. Schneeloch, G. D. Gu, F. Wang and N. L. Wang, Magnetoinfrared spectroscopy of landau levels and Zeeman splitting of three-dimensional massless Dirac fermions in $ZrTe_5$, *Phys. Rev. Lett.* **115**, 176404 (2015).

[32] D. Xiao, M.-C. Chang and Q. Niu. Berry phase effects on electronic properties, *Rev. Mod. Phys.* **82**, 1959 (2010).

[33] See Supplemental Materials for temperature dependence of Hall effect, angluar dependence of magnetoresistance with magnetic field rotating in the ab plane and the details of the calculations of anomalous Hall conductivity and longitudinal negative magnetoresistance.

**Figures and Captions**

**Figure 1**

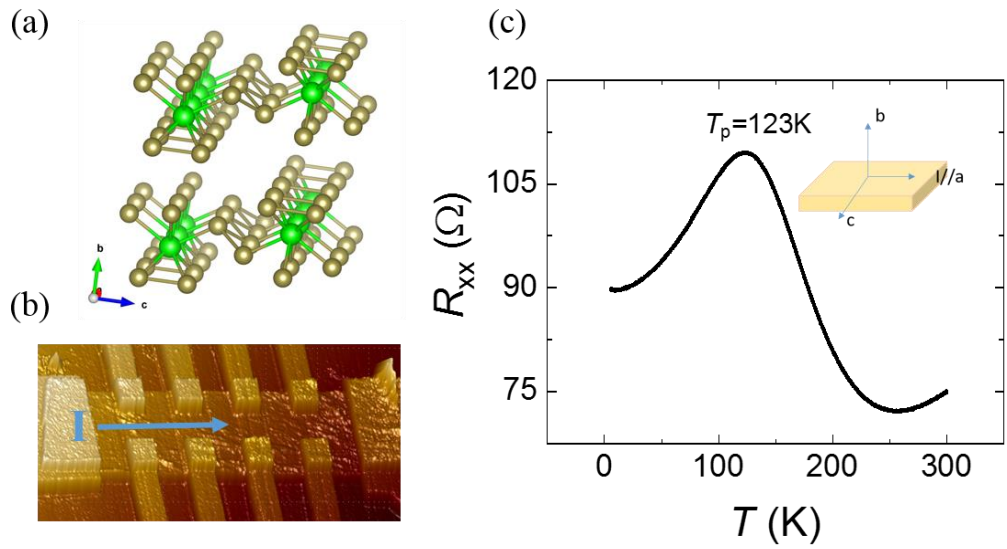

Figure 1. Illustration and characterization of device. (a) Atomic structural schematics of $ZrTe_5$. Here green and grey balls denote zirconium and tellurium atoms, respectively. (b) 3D AFM image of a $ZrTe_5$ device with thickness of 42 nm. The blue arrow indicates the direction of electrical current. (c) $R$-$T$ curve of the device at zero magnetic field. Inset: crystal orientation and current of the $ZrTe_5$ device.

**Figure 2**

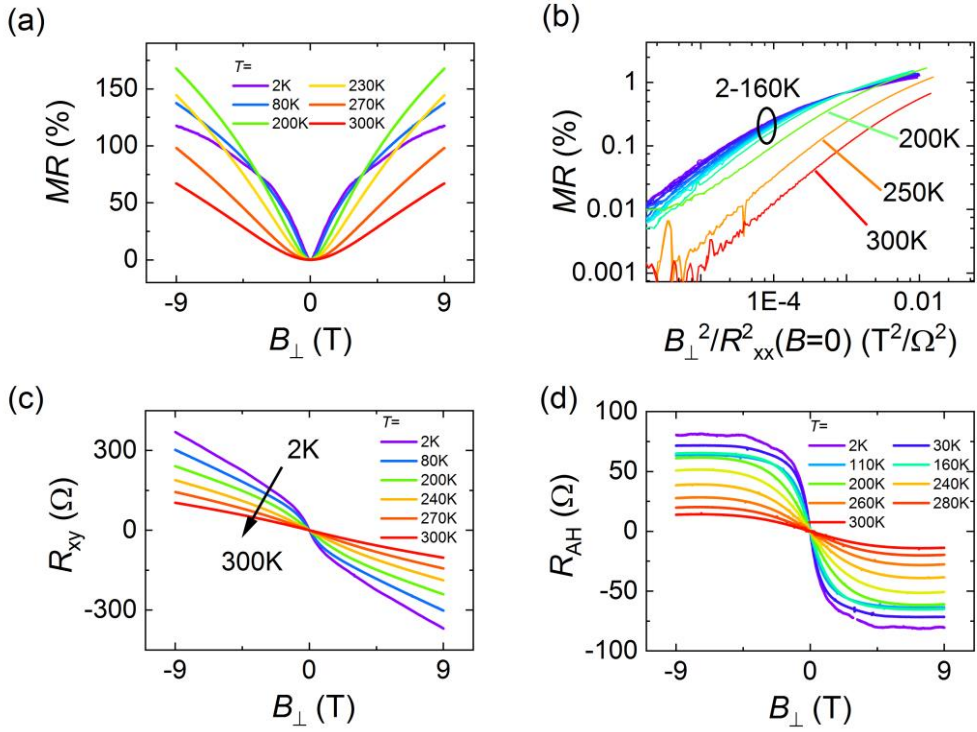

Figure 2. Magnetoresistance and Hall effect with perpendicular magnetic field. (a) *MR* versus perpendicular magnetic field at different temperatures. (b) The Kohler's plots of the *MR* curves at temperatures ranging from 10K to 300K. (c) Hall resistance at different temperatures. (d) Hall anomaly at different temperatures which shows saturation at high magnetic field.

**Figure 3**

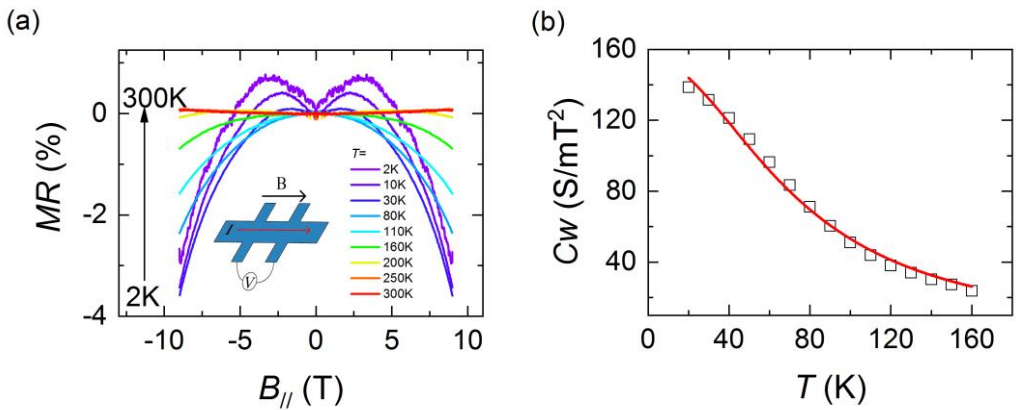

Figure 3. Chiral anomaly with parallel magnetic field. (a) Magnetoresistance vs. in-plane magnetic field parallel to the direction of current at different temperatures. Inset: measurement configuration of the *NMR*. (b) Temperature dependence of coefficient $C_w$ in the magneto-conductance due to chiral anomaly.

**Figure 4**

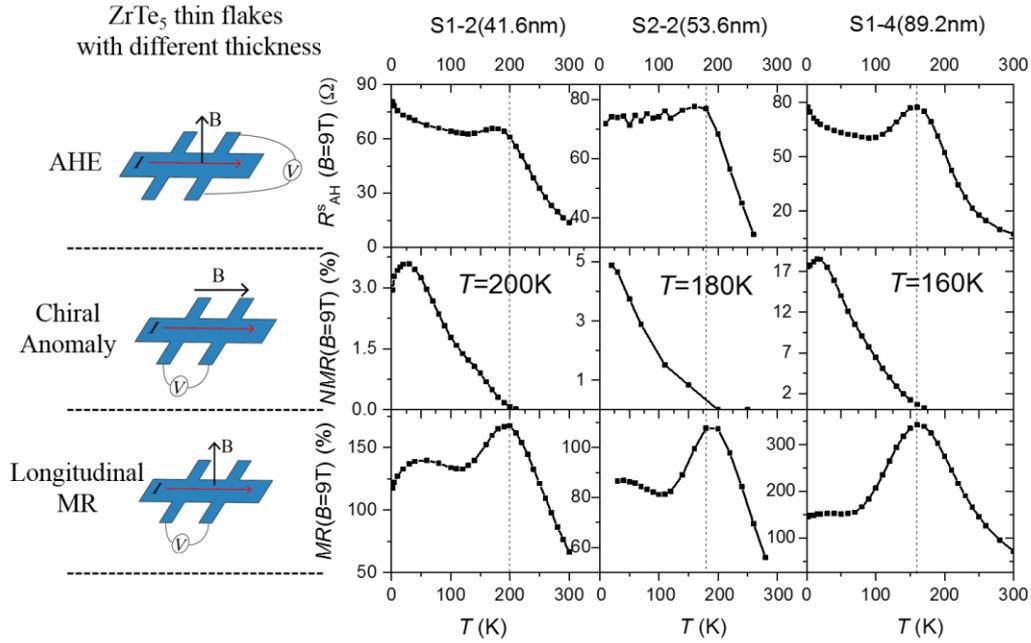

Figure 4. Summary of temperature dependent *MR* at different measurement configurations. The upper panel shows the saturated Hall anomaly $R_{AH}^s(B=9\text{T})$ with out-of-plane magnetic field. The middle panel represents the amplitude of negative magneto-resistance $NMR(B=9\text{T})$ with magnetic field parallel to the direction of current. The lower panel depicts the longitudinal magnetoresistance $MR(B=9\text{T})$ with out-of-plane magnetic field.

**Figure 5**

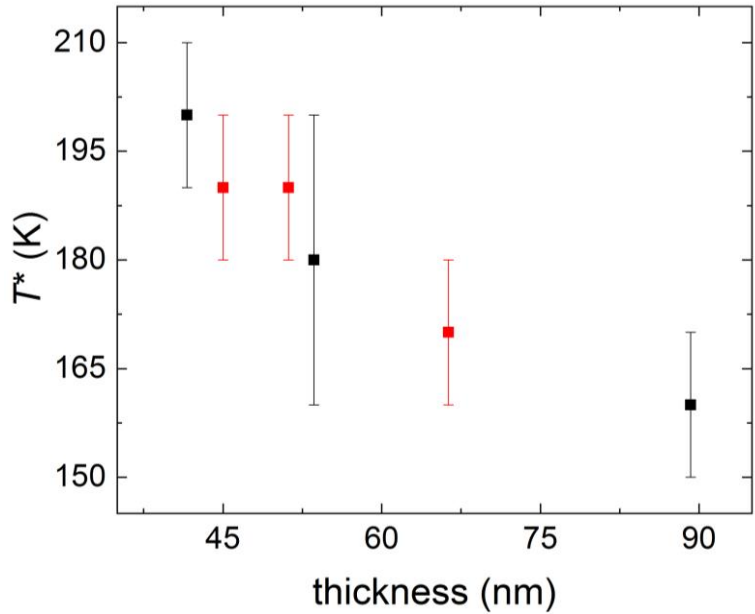

Figure 5. The dependence of $T^*$ on ZrTe$_5$ sample thickness. $T^*$ decreases monotonically from 200K to 160K with increasing thickness of the ZrTe$_5$ thin flakes from 42nm to 89nm. Here black squares are measured from crystals purchased from PrMat Technology company [www.prmat.com] and red squares are measured from crystals obtained from Brookhaven National Lab [19].

**Figure 6**

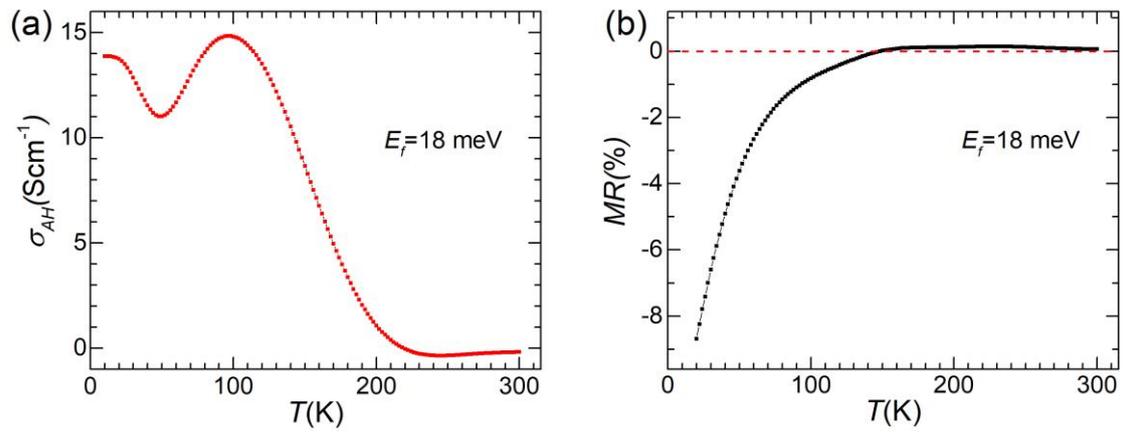

Figure 6. Calculated temperature dependence of anomalous Hall conductivity (a) and longitudinal negative magnetoresistance (b). Both physical quantities tend to zero around 200 K. The results are based on the same parameters, and with Fermi energy at 18 meV and magnetic field at 9 T.